# FPGA implementation of a deep learning algorithm for real-time signal reconstruction in radiation detectors under high pile-up conditions

**J. L. Ortiz Arciniega,**[a]  **F. Carrió**[b] **and A. Valero**[b]

[a] *Universitat de València,*
  *Avinguda de l'Universitat s/n, Country, Burjassot, Spain.*

[b] *Instituto de Física Corpuscular (Universitat de València-CSIC),*
  *Catedrático José Beltrán 2, Paterna , Spain.*
  *E-mail*: orarjo@alumni.uv.es

ABSTRACT: The analog signals generated in the read-out electronics of radiation detectors are shaped prior to the digitization in order to improve the signal to noise ratio (SNR). The real amplitude of the analog signal is then obtained using digital filters, which provides information about the energy deposited in the detector. The classical digital filters have a good performance in ideal situations with Gaussian electronic noise and no pulse shape distortion. However, high-energy particle colliders, such as the Large Hadron Collider (LHC) at CERN, can produce multiple simultaneous radiation events, which produce signal pileup. The performance of classical digital filters deteriorates in these conditions since the signal pulse shape gets distorted. In addition, this type of experiments produces a high rate of collisions, which requires high throughput data acquisitions systems. In order to cope with these harsh requirements, new read-out electronics systems are based on high-performance FPGAs, which permit the utilization of more advanced real-time signal reconstruction algorithms. In this paper, a deep learning method is proposed for real-time signal reconstruction in high pileup radiation detectors. The performance of the new method has been studied using simulated data and the results are compared with a classical FIR filter method. In particular, the signals and FIR filter used in the ATLAS Tile Calorimeter are used as benchmark. The implementation, resources usage and performance of the proposed Neural Network algorithm in FPGA are also presented.

Contents

## 1. Introduction

Radiation detectors are used to study the shower of particles produced in high-energy particle colliders. The particles crossing the detectors produce signals that are gathered by the data acquisition electronics to measure their track, momentum, energy or time.

Calorimeters measure the energy of the particles by absorbing completely their energy [1]. The read-out electronics produce analog pulses proportional to the energy deposited, which are conditioned and shaped prior to the digitization in order to improve the signal to noise ratio (SNR). The real amplitude of the analog signal is then obtained using digital filters, which provides information about the energy deposited in the detector. The classical digital filters have a good performance in ideal situations with Gaussian electronic noise and no pulse shape distortion. Due to its simplicity, they can be executed in real time in sequential Digital Signal Processors (DSPs) in relatively low rate experiments without introducing dead-time.

However, high-energy particle colliders, such as the Large Hadron Collider (LHC) [2] at CERN, can produce multiple simultaneous radiation events, which produce high event rates and signal pileup. The performance of classical digital filters deteriorates in these conditions since the signal pulse shape gets distorted. In addition, this type of experiments produces a high rate of collisions requiring high throughput and low latency data acquisitions systems.

In order to cope with these harsh requirements, new read-out electronics systems are based on high-performance FPGAs, which permit the utilization of more advanced real-time signal reconstruction algorithms to exploit the parallelization capabilities of FPGAs.

A deep learning method has been studied for real-time signal reconstruction in high pileup and high event rate radiation detectors. The algorithm is trained offline using simulated data, which emulates the signal conditions in terms of noise and pileup of the experiment. Then, the trained network is loaded into the FPGA for real time operation.

## 2. Classical signal reconstruction algorithms

In classical Calorimeters the signals produced in the detector are conditioned, shaped and digitized in the front-end electronics and transmitted to the back-end for signal reconstruction. Then, digital filters are used to obtain the real amplitude of the analog pulse, which is proportional to the energy deposited in the detector. Figure 1 shows a typical pulse shape and the digital samples generated in the ATLAS Tile Hadronic Calorimeter [3] in the LHC at CERN. A digital FIR filter called Optimal Filtering [4] is used to obtain the amplitude, phase and baseline (pedestal) of the pulse. The Optimal Filtering method is a FIR filter which exploits the knowledge of the pulse shape of the electronics and the amount of expected pileup to reduce the contribution of noise and determine the time of deposition. Thus, any pulse shape distortion strongly affects the performance of the Optimal Filtering method. In particular, signal pileup different than expected deforms the signal of interests and biases the results of the signal reconstruction. This algorithm is sequentially executed in Digital Signal Processors (DSPs) to reconstruct the signals for 48 channels. The result must be provided in less than 10 μs determined by the event rate not to introduce dead-time in the detector.

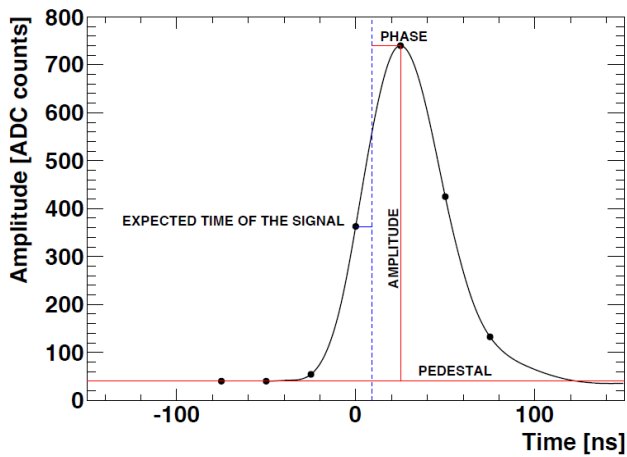

**Figure 1.** Analog pulse shape with the associated digital samples and the reconstructed magnitudes.

### 2.1. Signal reconstruction in high pileup and high rate environments

The performance of the digital filters explained in the previous section significantly deteriorates in the presence of pulse pileup since they rely in the knowledge of the pulse shape.

Moreover, an increase in the event rate is also assumed which would require a reduction of the algorithm latency.

This means that a new signal reconstruction algorithm capable of supporting large signals pileup and with low latency is required to operate under these conditions. A new algorithm based on supervised learning, specifically on Artificial Neural Networks (ANN) has been studied.

The proposed method has been implemented to use 9 digital samples, which shows the most optimal utilization of FPGA resources in terms of performance. In addition, the method has been implemented to reconstruct up to 48 channels in parallel in one single FPGA in order to use the TileCal example as a benchmark. This method provides reduced and deterministic latency, thanks to the characteristics of the ANN that allow great interconnection and parallelism.

## 3. Neural Network algorithm for signal reconstruction

### 3.1. Neural Network model and architecture

For the design of the ANN, the neuronal model proposed by McCulloch and Pitts in 1943 [5] has been used, which despite being very old is still the most extended for the different architectures of ANNs. In this model, whose schema is shown in Figure 2, the output of the neuron represented by $y$ is obtained by applying an activation function $g$ to the weighted sum among the multiple input signals $\{x_1, x_2, ..., x_n\}$ and their corresponding synaptic weights $\{w_1, w_2, ..., w_n\} \pm$ a bias $\theta$ that allows us to specify an appropriate threshold. The following expressions synthesize the result produced by the artificial neuronal model proposed by McCulloch and Pitts:

$$u = \sum_{i=1}^{n} w_i \cdot x_i \pm \theta$$

$$y = g(u)$$

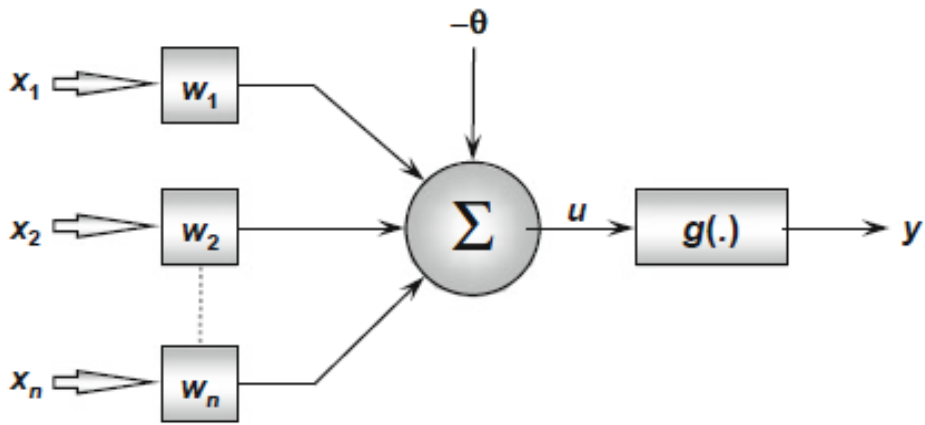

**Figure 2**. Representation of an artificial neuron.

The most optimal activation function in complex nonlinear systems is the sigmoidal, because it allows gradual processing responses where the output results are real numbers between 0 and 1. In this work, a variant of this function is used, due to the need for negative results given to the characteristics of the system. Its mathematical expression is as follows:

$$g(u) = \frac{2}{1 + e^{-2 \cdot u}} - 1$$

A multilayer feedforward architecture is used, since it is optimized in the identification of systems or signals, where the neurons are organized forming 3 different types of layers: input layer, hidden layers and output layer. Another characteristic of this type of architecture is that the learning algorithm follows a supervised training, which means that synaptic weights and biases of the network are calculated from a set of inputs and desired outputs (targets).

The design of the ANN is done through the Neural Network Fitting application of MATLAB [6], which allows developing a multilayer feedforward network and generates the corresponding code for its subsequent optimization. The first step corresponds to selecting the inputs and targets of the network. The dataset used both for training and to evaluate the performance of the algorithms is generated using a pulse simulator. The simulator generates pseudorandom pulse amplitudes simulating the signals produced by the energy deposited by particles crossing the detector (true values and targets) and then produces a byte-stream emulating the digital samples produced by the electronics. Following this, the dataset is divided into training, validation and test subsets to proceed to select the number of hidden layers and train the network.

Three different algorithms are used to train the network and their performance is evaluated comparing the Mean Square Error (MSE), which must be close to 0, and the correlation between the target and the obtained result (R), where values are sought close to 1. This comparison, presented in Table 1, provides an optimal view of which algorithm the network should be trained, the ANN is composed of a single hidden layer.

| Comparison of training algorithms | | |
|---|---|---|
| **Algorithms** | **MSE** | **R** |
| **Levenberg-Marquardt** | -0.5442 | 0.9875 |
| **Bayesian Regularization** | -0.4615 | 0.9875 |
| **Scaled Conjugate Gradient** | 4.3820 | 0.9874 |

**Table 1.** Comparison of training algorithms.

It can be concluded that the most optimal result is obtained by using the Bayesian Regularization training algorithm.

### 3.2. Implementation of the Neural Network algorithm in FPGA

The code corresponding to a feedforward network with 1 hidden layer, trained with the Bayesian Regularization algorithm and generated by the Neural Network Fitting application, describes the sequential steps carried out by the ANN. This includes three functions to normalize the inputs, activate the hidden layer by means of a variant of the sigmoidal function and denormalize of the outputs. In addition, the code includes some constants called "Neural Network Constants" that correspond to synaptic weights and network biases.

The implementation of the algorithm is done for the Xilinx Virtex-7 XC7VX485T FPGA, which is the model used for the upgrade of the TileCal readout electronics [7]. The first step is to convert the code generated by the Neural Network Fitting application to synthesizable VHDL and Xilinx Series 7 FPGA cores. Using the HDL Coder application, also from MATLAB, the functions and operations using floating point arithmetic are identified, and then optimized to fixed-point logic in order to obtain synthesizable code.

The normalization and denormalization functions are simplified to basic mathematical operations, since they only perform addition, subtraction and multiplication using the "Neural Network Constants". On the other hand, the variant of the sigmoidal function (activation function) performs complex operations, as divisions and exponentials, and they are implemented as Look-Up Tables (LUT) containing the necessary values not to lose resolution.

The designed LUT is a vector that contains all the possible values of the activation function within a range of *[-1, 1.2]*. The resolution of the algorithm was studied for different LUT sizes in order to optimize the memory resources used in the FPGA to store the LUT values. Table 2 shows the regression and RMSE for different sizes of the LUT. The most optimal result was obtained for 5,000 samples which presents a correlation between the true and the reconstructed result of 98.79%, being only 0.03% lower than a LUT with 20,000 samples.

| Number of Samples | RMSE | Regression (%) |
|---|---|---|
| **20,000 (Original algorithm)** | 45.9167 | 98.829 |
| **10,000** | 45.8876 | 98.813 |
| **5,000** | 46.0348 | 98.799 |
| **1,000** | 48.4637 | 98.721 |
| 500 | 51.4995 | 97.712 |
| 100 | 57.80 | 96.803 |

**Table 2.** Comparison of different sizes of LUT.

The fixed-point algorithm using the LUT for the activation function is implemented for the Virtex-7 FPGA including additional data flow and control blocks to obtain the final specifications:
- A digital pulse must be reconstructed every 25 ns (40 MHz).
- Each reconstructed pulse is composed by the last 9 samples received which are obtained with a 9-position shift register.
- The inputs are normalized and sent to the first layer of the network (hidden layer).
- In the hidden layer, the mathematical operations corresponding to the neuronal model of McCulloch and Pitts are performed and the output of the layer is activated using the LUT [7].
- The output of the hidden layer serves as input to the second layer of the network (output layer), where a process similar to the previous layer is performed. In this case, the output is denormalized to obtain the initial scale.
- An output result is obtained after one clock period, so that the latency is minimal and deterministic.

Figure 3 shows the process carried out by the Artificial Neural Network through its two layers:

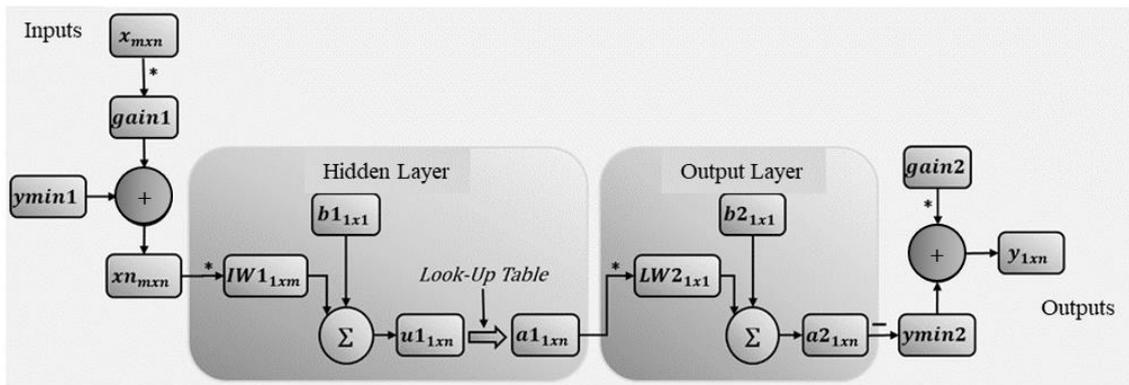

**Figure 3.** Block Diagram of the ANN.

## 4. Results

The performance of the proposed ANN algorithm has been studied using a data simulator, which permits the comparison between the obtained result and the true input. The TileCal Optimal Filtering algorithm is used as a figure of merit to evaluate the performance of the proposed method. The study has been done in MATLAB using the optimized fixed-point arithmetic version of the algorithm. Finally, the analysis has been replicated in the FPGA to validate the implementation and to qualify the migration of the algorithm to FPGA resources.

### 4.1. Data simulator

In order to verify the correct functioning of the algorithm in MATLAB, it is necessary to develop a signal data simulator. The simulator generates pseudorandom pulse amplitudes following a normalized Gaussian distribution covering the entire amplitude range. For every input amplitude the simulator generates the corresponding digital samples emulating the pulse shaping of the electronics. For this study the TileCal pulse shape has been used in order to use the Optimal Filtering reconstruction result as reference. In the real detector, the possibility of having energy deposited in a channel in two consecutive events (pileup) is proportional to the accelerator instantaneous luminosity. The data simulator has been implemented to emulate the harshest possible pileup conditions, thus generating a non-zero energy deposition in every bunch crossing.

The study has been done using a generated input data of 20,000 events emulating the same number of consecutive collisions in the detector. Then, each event has been reconstructed using both the Optimal Filtering and the ANN methods to obtain the corresponding 20,000 results. The Optimal Filtering weights have been calculated for the simulated pileup conditions.

Figures 4 and 5 shows the linear regression between the targets (input data) and the result for the ANN and Optimal Filtering methods. The correlation for the ANN is 97.73% whereas for the Optimal Filtering algorithm it is 90.29%.

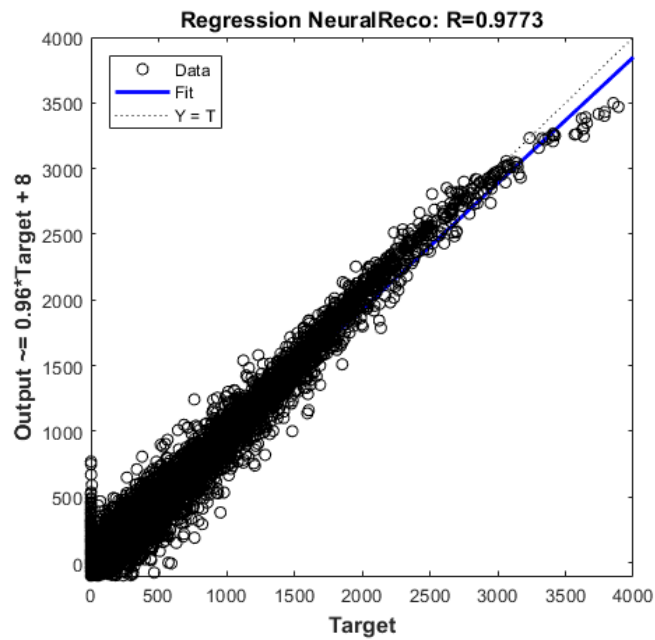

**Figure 4.** Correlation between the output of the ANN algorithm and the targets.

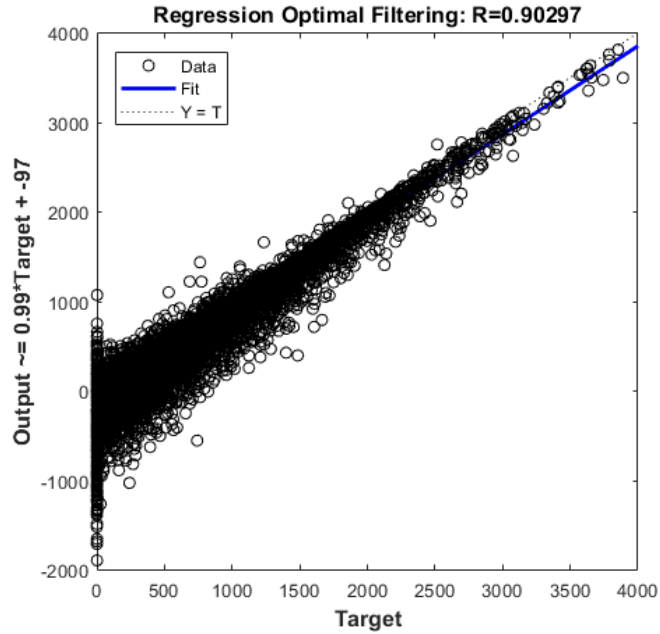

**Figure 5.** Correlation between the output of the OF algorithm and the targets.

### 4.2. Signal reconstruction results in FPGA

The same data set used for the performance study in MATLAB has been used to evaluate the behavior of the algorithm in the FPGA. The data set has been introduced in an internal FPGA memory and then the reception of the digital samples at 40 MHz has been simulated. The code has been implemented for one channel and then scaled for 48 channels, which represents one TileCal detector module.

The results obtained in the FPGA for the 20,000 events in the data set match exactly the results obtained in MATLAB with the fixed-point arithmetic version of the algorithm. Therefore, the implementation of the proposed algorithm in FPGA does not present any limitation. In addition, the number of FPGA resources used is very reduced except for the number of DSP slices, where the bulk of the processing is done (Table 3). Thus, the parallelization could be increased even further to process more than 48 channels in parallel.

| Resource summary | Available | Utilization | |
|---|---|---|---|
| | | ANN | OF |
| Slice registers | 607,200 | 695 | 98 |
| Slice LUTs | 303,600 | 1297 | 87 |
| RAM blocks | 1,030 | 4 | 0 |
| DSPs | 2,800 | 13 | 3 |
| Timing summary | | | |
| Latency | - | 5 clk cycles | 1 clk cycles |
| Minimum period | - | 6.047 ns | 1.378 ns |
| Maximum frequency | - | 165.375 MHz | 725.689 MHz |

**Table 3.** Comparison of the resources used by the FPGA.

## 5. Conclusions

An algorithm based on neural networks has been proposed for real-time signal reconstruction in high pileup and high rate radiation detectors. The performance of the new method has been studied using simulated data that emulates the harshest possible pileup conditions on a detector. The Optimal Filtering digital filter used in the ATLAS Tile Calorimeter has been used as benchmark to evaluate the performance and behavior of the Neural Network method.

It is important to note that the two methods have been evaluated in a free-running mode where it is assumed that non-zero signal is present in every event. The ANN method is trained to detect real and fake energy depositions in the digital signals whereas the Optimal Filtering would require additional logic to implement this detection.

The obtained results show better performance of the new Neural Network compared with the Optimal Filtering method. In particular, the regression obtained for the Neural Network method is 0.987 while it is 0.903 for the Optimal Filtering. The utilization of FPGA resources for the ANN method is higher than for Optimal Filtering case but in both cases, it represents a small percentage of the total resources available in the FPGA used.

The latency of the ANN is 5 clock cycles while it is just one cycle for the Optimal Filtering. The method is implemented in pipeline mode, so it can reconstruct a signal at the maximum operation frequency of 165 MHz. On the contrary, the Optimal Filtering running in DSPs has a higher operation frequency (725 MHz) but effectively can reconstruct at a maximum event rate of 100 kHz due to the sequential operation of the DSP.